\documentclass[12pt]{iopart}

\usepackage{iopams}
\usepackage{graphicx}
\usepackage[latin1]{inputenc}
\usepackage{amssymb}

\DeclareGraphicsExtensions{.eps,.ps,.eps.gz,.ps.gz,.eps.Z,.pdf,{}}

\begin{document}

\title{Molecular vibrational cooling by Optical Pumping with  shaped femtosecond pulses}
\author{Dimitris Sofikitis$^{1}$, Sebastien Weber$^{2}$, Andrea Fioretti$^{1}$, Ridha Horchani$^{1}$%
, Maria Allegrini$^{1,3}$, B\'eatrice Chatel$^{2}$, Daniel Comparat$^{1}$ and Pierre Pillet$^{1}$ \\
$^{1}$ Laboratoire Aim\'{e} Cotton, CNRS, Universit\'{e} Paris-Sud, B\^{a}t. 505, 91405 Orsay, France.\\
$^{2}$ Laboratoire Collisions, Agr\'egats, R\'eactivit\'e
(UMR 5589, CNRS - Universit\'e Paul Sabatier Toulouse 3), IRSAMC,
Toulouse, France
\\
$^{3}$ CNISM, Dipartimento di Fisica, Universit\`{a} di Pisa, Largo Pontecorvo, 3 56127 PISA, Italy.\\
E-mail: daniel.comparat@lac.u-psud.fr}
 
\begin{abstract} 
Some of us have recently reported in Science~\protect{\cite{2008Sci...321..232V}} vibrational cooling of translationally cold Cs$_2$ molecules into the lowest vibrational level $v=0$ of the singlet $X ^{1}\Sigma_{g}$ ground electronic state. Starting from a sample of cold molecules produced in a collection of vibrational levels of the ground state, our method was based on repeated optical pumping by laser light with a spectrum broad enough to excite all populated vibrational levels but frequency-limited in such a way to eliminate  transitions from $v=0$ level, in which molecules accumulate. In this paper this method is generalized to accumulate molecules into an arbitrary selected "target" vibrational level. It is implemented by using ultrashort pulse shaping techniques based on
Liquid Crystal spatial light modulator. In particular  a large fraction of the initially present molecule is transferred into a selected vibrational level such as $v=1$, 2 and 7. Limitations of the method as well as the possible extension to rotational cooling are also discussed.
\end{abstract}


\section{Introduction}

The manipulation of atomic or molecular quantum dynamics and the availability of robust and selective methods of executing
population transfer in quantum systems is essential for a variety of fields.  We could mention precision spectroscopy, quantum computing, 
control of molecular dynamics and chemical reactions, biophotonics, nanoscience or production of cold
molecules~\cite{Ye2005, Shapiro2003, Alessandro2007, 2000Sci...288..824R}. 
In particular, the important activities developed in the cold molecule domain through precise control of both internal and external 
degrees of freedom of a molecule is expected to lead to significant advances in collision dynamics of  chemical reactions, molecular 
spectroscopy, molecular clocks, fundamental test in physics, controlled photo-chemistry studies, and also in quantum
computation with the use of polar molecules~\cite{2004EPJD...31..149D,Krems,hutson-2006-25,DulieuJPB2006,2008PCCP...10.4079K,Smith2008}.
Several theoretical approaches have been proposed to control the internal degrees of freedom of a cold molecule such as the use of an 
external cavity to favor spontaneous emission toward the lowest ro-vibrational level~\cite{2007PhRvL..99g3001M} or the controlled 
interplay of coherent laser fields and spontaneous emission 
through quantum interferences between different
transitions~\cite{1993JChPh..99..196B,1999FaraDisc,2001CP....267..195B,2001PhRvA..63a3407S}. Finally, the use of a tailored incoherent broadband light 
source for internal cooling of molecule has been suggested~\cite{2002PhRvL..89q3003V,2004JPhB...37.4571V}.

During the last two decades many results including coherent control
\cite{Assion98,MonmayrantCT-spirograph-06, Motzkus02bio},
compression of optical pulses \cite{efimovGA2000} and optical
communications \cite{weinerjlt98} have been obtained by the use of
arbitrarily shaped optical waveforms. Most of these works were
spurred by technological breakthroughs. These pulse shaping
techniques have been reviewed in details \cite{2000RScI...71.1929W}. Due to
their ultrashort duration, femtosecond pulses are not easily shaped
in the time domain. Thanks to the Fourier transformation, the common
way to synthetize them is in the spectral domain. The most usual
device for both high fidelity and wide flexibility of shapes
involves a pair of diffraction gratings and lenses arranged in a
zero-dispersion line \cite{Martinez87} with a pulse shaping mask at
the Fourier plane. In this paper this technique will be used to
improve the vibrational cooling of molecules using amplitude shaping
only.

Some of us have recently published an experimental realization of the  vibrational cooling  based on  optical pumping using a train of several identical weak femtosecond laser pulses~\cite{2008Sci...321..232V}. Cs$_2$ molecules initially formed via photoassociation of cold Cs atoms in several vibrational levels, $v$, of the electronic ground state were
redistributed 
in the ground state via a few electronic excitation - spontaneous emission cycles
by applying    
 a femtosecond broadband laser. The laser
pulses were shaped to remove the excitation frequency band of the $v=0$ level, preventing excitation from that state and leading to efficient accumulation in the lowest vibrational level of the singlet electronic state. 

Here, using  the flexibility of femtosecond pulse shaping
techniques~\cite{2000RScI...71.1929W}, this incoherent population pumping method is extended in order to accumulate molecules into other single selected vibrational level than the sole $v=0$ one.
The outlook of this paper is the following: we first recall the principle for transferring populations from several energy eigenstates
into the lowest vibrational level. We then briefly describe then our experimental apparatus in its main parts: the magneto-optical trap where the cold molecules are produced and the pulse shaping apparatus based on a Liquid crystal spatial light modulator placed in
the Fourier plane of a highly dispersive 4f line \cite{pulseshaperRSI04}. Then we report our new experimental result: the selective vibrational cooling of the molecules into one given vibrational level, on demand. Examples are reported for $v=0$, $v=1$, $v=2$ and $v=7$. In order to improve the efficiency of the optical pumping, we experimentally investigate in more details the cooling into $v=1$. Finally, perspectives for very large band vibrational cooling and for rotational cooling are briefly theoretically addressed.

\section{Broadband laser cooling of the molecular vibration}

\subsection{Optical pumping} 

The main idea, in the optical pumping, as performed in Ref. \cite{2008Sci...321..232V}, is to use a broadband laser tuned to
the transitions between the different vibrational levels, which we label $v_{\rm{X}}$ and $v_{\rm{B}}$, belonging respectively to the singlet-ground-state $X^1\Sigma_{\rm g}$, hereafter simply referred as $X$, and to an electronically excited state, the ${\rm B}^{1}\Pi_{u}$ state of the Cs$_2$ system, hereafter referred as $B$. 
 The goal is to start from a given vibrational distribution of  $v_{\rm X}$ values and to transfer 
 it into a single target $v_{\rm X}$ level.
 The absorption - spontaneous emission cycles lead, through optical pumping, to a redistribution of the vibrational population into the ground state according to the scheme:
\begin{equation}
	{\rm Cs}_{2}(v_{\rm X})+h\nu_{\rm las} \longrightarrow {\rm Cs}_{2}(v_{\rm B}) \longrightarrow {\rm Cs}_{2}(v^{'}_{\rm X}) 
	+h\nu_{\rm em}.
\label{equ:Equ1}
\end{equation}
where $h\nu_{\rm las}$ and $h\nu_{\rm em}$ are the energies of the laser and of the spontaneously emitted photons respectively.
The broadband character of the laser permits repetition of the pumping process from multiple vibrational $v_{\rm X}$ levels. By removing the laser frequencies corresponding to the excitation of a selected $v_{\rm X}$ level, we make it impossible to pump molecules out of this level, thus making $v_{\rm X}$ a dark state. As time progresses a series of absorption - spontaneous emission cycles described by Eq~(\ref{equ:Equ1}), leads to an accumulation of the molecules in the $v_{\rm X}$ level.

In the experiment with cold cesium dimers reported in Ref. \cite{2008Sci...321..232V}, 
the starting  given vibrational distribution was $v_{\rm X} = 1-10$, the target level was $v_{\rm X} = 0$, the  broadband laser was a
 Ti:Saphire femtosecond mode-locked laser (standard deviation-gaussian bandwidth 54\ cm$^{-1}$, average intensity of $50$ mW/cm$^{2}$) and the shaping was a simple cut of the blue part of the laser spectrum, which otherwise would have been able to excite the $v_{\rm X} = 0$ level. In the work presented here we  extend this incoherent depopulation pumping method by using a high resolution pulse shaper. The results are described in the following sections.

\subsection{Cold molecule production and pulse shaping}
\label{MOTandSLM}

As in the work presented in Ref. \cite{2008Sci...321..232V},
the cold molecule formation 
 is achieved in a Cs vapor-loaded Magneto-Optical Trap (MOT) via 
photoassociation where  two  atoms
resonantly absorb a photon  to
create a molecule in an excited electronic state which decay  into stable 
deeply bound vibrational levels of the singlet molecular ground $X$ state. Photoassociation is achieved using a cw Titanium:Sapphire 
laser (intensity 300 W$/$cm$^2$) pumped by an Argon-ion laser. 

The stable molecules are then ionized by 
Resonantly Enhanced Multiphoton Ionization (REMPI) with the excited ${\rm C}^{1}\Pi_{u}$ molecular 
state as intermediate state. The REMPI detection uses a pulsed dye laser (wavenumber $\sim$16000 
cm$^{-1}$, spectral bandwidth 0.3 cm$^{-1}$) pumped by the second harmonic of a pulsed Nd:YAG laser (repetition rate $10\,$Hz,
duration 7ns). The formed Cs$_{2}^{+}$ ions are detected using a pair of microchannel plates through a time-of-flight mass 
spectrometer. By scanning the REMPI laser wavelength the experimental spectrum already presented in Ref.
\cite{2008Sci...321..232V} and
visible in Fig. \ref{fig2:v=0} d) is monitored. The 
observed lines represent transitions from  vibrational ground states  $v_{{\rm X}}=1-7$ level  to various levels $v_{{\rm C}}$ of the C state 
\cite{1982JChPh..76.4370R} (a more detailed study of the process is performed in Ref. \cite{2008PRA}). 
 The present 
low REMPI resolution does not provide the capability of analyzing the rotational population of the molecules.


In our experiment,
the pulse shaped femtosecond laser, used to achieve vibrational cooling, is provided by a Kerr-Lens-Mode-locking Ti: Sapphire oscillator
 with a repetition rate of 80~MHz (12.5\,ns between subsequent pulse). 
 The central wavelength is 773\,nm. The spectral
FWHM is around 10\,nm. 

\begin{figure}[ht]
\centering
\includegraphics*[width=0.85\textwidth]{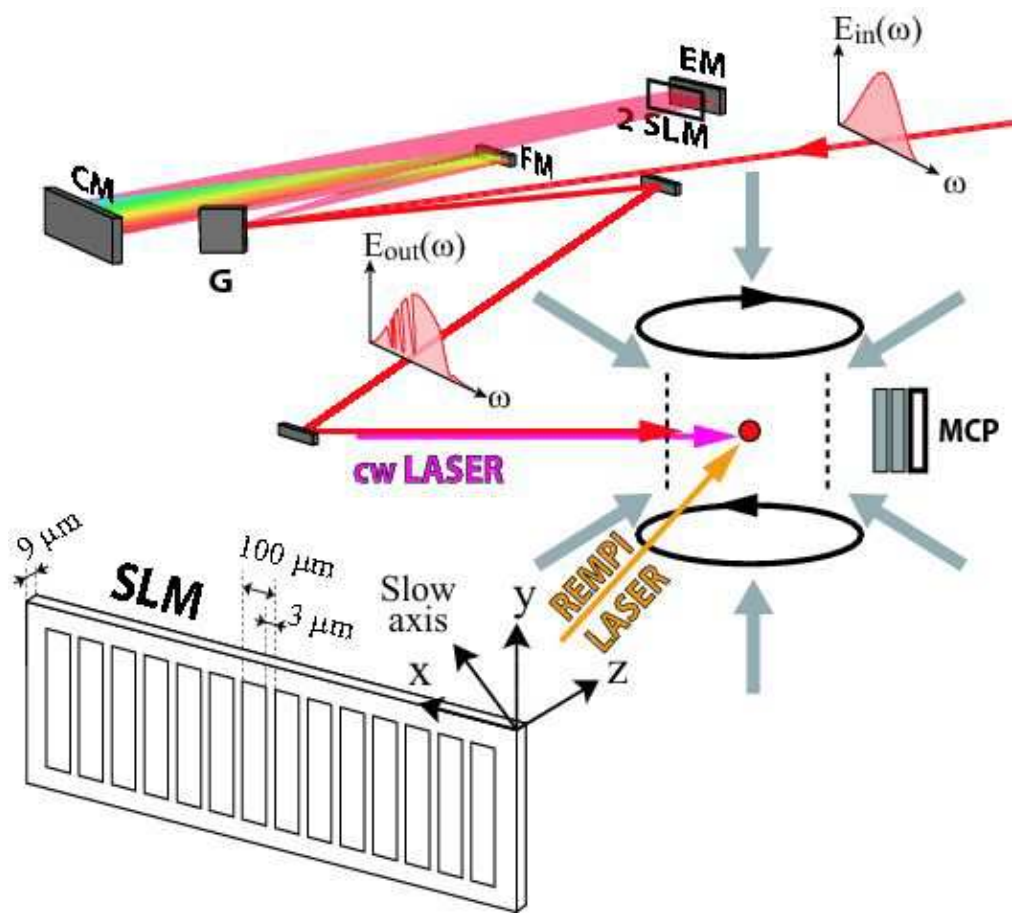}
 \caption{Experimental setup for the pulse shaper and the cold molecule production and detection. Upper part:  folded zero dispersion line . The beam is dispersed by the grating \textsf{G} and then each spectral component is spatially separated and  focalized by the cylindrical mirror \textsf{CM} in the Fourier plane. \textsf{FM} is a plane folded mirror. Both Spatial Light Modulator (SLM, detailed in the lower-left side of the figure) are at the Fourier plane. An end mirror \textsf{EM} is placed just after the SLM and the beam goes twice through the folded line. The shaped light is then send to the cold molecular cloud which is created by photoassociation (with the cw laser) of an atomic Cs cloud
cooled by a standard 6 beams vapor-loaded Magneto-Optical Trap (MOT). The molecules are then detected using a REMPI ionization laser creating Cs$_2^+$ ions, which are accelerated by an electric field created with the two grids surrounding the cloud, and monitored using Micro-Channel Plate (MCP).}
\label{shaper}
\end{figure}

In order to control the optical pumping,
the spectrum of this femtosecond laser
 is shaped by a high resolution pulse
 shaper \cite{pulseshaperRSI04}. This one is composed of a dual
 Liquid Crystal spatial light modulator (SLM) placed in the Fourier plane of a folded
 zero dispersion line (see Fig. \ref{shaper}) which allows phase and amplitude modulation
\cite{WeferNelson95slm}. 

Let us recall  some basics: The incoming laser
field $E_1$ is polarized along the $x$ axis, $z$ being the
propagation axis. The Liquid-Crystals are rodlike molecules that have a
variable birefringence. They tend to align themselves with an
applied electric field. In this set-up, the Liquid-Crystal molecules are aligned
along axis at 45 degrees (for the first SLM) and $-45$ degrees (for the second SLM) in the $x$-$y$ plane. The $x$
polarized contribution  of the outcoming field $E_2$, noted $ E_{2x}$, can be
written as

\begin{equation}
 E_{2x}(x)=E_{1}(x)\times
\exp{\left(i(\phi_1(x)+\phi_2(x))\right)}\cos\left( \phi_1(x)-\phi_2(x)\right)
\label{eqSLM_Bx}
\end{equation}
where $\phi_1$ and $\phi_2$ are the voltage dependent birefringences
of the first and the second Liquid Crystal array, respectively. $\phi_1$ and $\phi_2$ depend on the $x$ value, i.e. on the considered pixel. So by using an
output polarizer along the $x$ axis, the output phase (retardation)
and amplitude (attenuation) can be set independently by controlling
the sum and the difference of both Liquid Crystal birefringences,
respectively. Moreover, the use of orthogonal polarizations on each
Liquid Crystal limits multiple diffraction \cite{WeferNelson95slm}. In this
present experiment, amplitude-only modulation is used.
 The Liquid-Crystal (SLM-S640/12) produced by Jenoptik company, possesses 640 pixels and has
been described in ref \cite{stobrawashap01} (stripes of $97 {\,\rm
\mu m}\times 10 {\,\rm mm}$ separated by gaps of $3 \rm\, \mu m$).
The birefringence of each pixel is controlled by a voltage with a
dynamic range of 12 bits. The non-linear
relation between voltage, incident wavelength and birefringence is
carefully calibrated \cite{stobrawashap01}. The regions of Liquid-Crystal between
the patterned electrodes cannot be controlled and are referred to as
gaps. In these gap regions the Liquid-Crystals behave in a first approximation
as though there were zero applied voltage so that the filter for the
gap regions is assumed to be constant across the array. This  limits the off-on ratio (degree of extinction) of theoretically 20\,dB (99\% intensity extinction).  However, probably due to imperfect experimental polarization, one has 
 measured, using an OCEAN Optics HR 4000 spectrometer, only an off-on ratio of 3\% of the light intensity in the worse case. In the 
simulation of the experiment this 3\% conservative value is the reference.

To avoid chromatic as well as off-axis aberrations, the setup shown in Figure \ref{shaper} is chosen.  The beam is first dispersed
by a gold-coated grating with 2000\,lines/mm and then focused in the
horizontal plane by a cylindrical mirror with a focal length of
600\,mm in the Fourier plane. The two Liquid-Crystal- SLMs (64 mm width) are
placed in the Fourier plane just behind the end mirror which
allows to fold the line without any misalignment effect thanks to
the large Rayleigth length. In this design, the beam passes twice through
the dual Liquid Crystal. By construction this setup provides a perfect symmetry
of the zero-dispersion line which  greatly simplifies  the alignment
procedure. The central wavelength of the 4f-line is set to 773\,nm.

These characteristics provide an average resolution of 0.06\,nm/pixel corresponding
to a spectral width of 38\,nm.
 This spectral width is large enough to transmit the spectral pedestal width of our
laser source (Full Width at Half Maximum (FWHM) of 10\,nm). The
sagital beam FWHM in the Fourier plane (57\,$\mu$m,
corresponding to an input beam diameter of 2.3\,mm) is roughly set
to the width of a pixel, therefore maximise the resolution of the
pulse shaper. Temporal replica inherent to these Liquid-Crystal devices are not a
limitation for this experiment.

The overall transmission in intensity of all the
device is currently of $60\%$, mainly limited by the grating's
efficiency which is enough in the present experiment because one use an average laser power of only few milliwatt focused on the molecular cloud with a waist of nearly $500\,\mu$m. A similar laser power of
3\,mW, corresponding to an intensity of $700\,$mW/cm$^2$, is used in the simulations.

\section{Experimental results}

\subsection{Vibrational cooling to the vibrational ground state}

\begin{figure}[htbp]
\begin{center}
\includegraphics[angle=-90,width=0.85\textwidth,keepaspectratio=true,clip=true,bb=10mm 21mm 193mm 270mm]{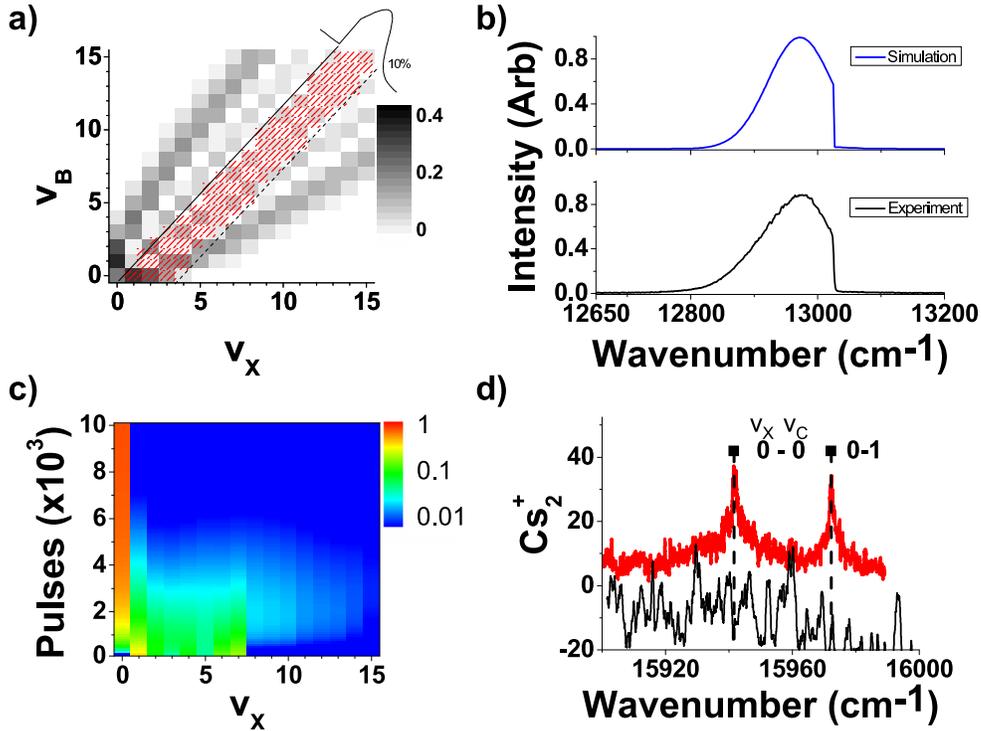}
\caption{Simulation and experimental results for the transfer in the vibrational ground state. (a) FC coefficients (grey scale) for transition between the X ground state and the B excited state. The hatched (red) area represents the transitions that are excited  by the pulse shaped laser intensity, only transitions with laser intensity more than 10\% of maximum intensity are shown. The sharp cut in the laser spectrum is represented by a solid line. (b) The shaped pulse used for the simulation (upper part) and in the experimental one (lower part). (c) Results of the simulation of the vibrational cooling where the (log scale) color level indicates molecular population. The accumulated population in each $v_{\rm X}$ level is plotted as a function of the number of femtosecond pulses. The femtosecond laser pulses occur
every $12.5\, $ns. After $10^4$ pulses the population in the $v_{\rm X} = 0$ level is 70\%. (d) REMPI molecular ion spectrum.  With the shaped femtosecond pulses (in red). This is the signature of $v_{\rm X} = 0$ molecules because only transition from $v_{\rm X} = 0$ to  $v_{\rm C} = 0$  and $v_{\rm C} = 1$ are present in the spectrum. The spectrum (in black, with an offset of -40 for clarity) without femtosecond pulses~\protect{\cite{2008Sci...321..232V}}, indicating the presence of molecules in several low vibrational levels, is reported for reference.}
\label{fig2:v=0}
\end{center}
\end{figure}

As already explained, accumulation in a given vibrational level comes from the fact that all the frequencies that could excite molecules decaying in this level during the optical pumping procedure are removed, making it a dark state of the system. 
	The  femtosecond  laser spectrum needed to realize such dark state is first theoretically calculated, and then implemented using the pulse shaper. 
	In order to predict the best possible laser spectrum,
	we have modeled the optical pumping in a very simple way. Using the known $X$ and $B$  potential curves and their rotational constants~\cite{1985JChPh..82.5354W, 1989CPL...164..419D}, we calculate the ro-vibrational energy levels as well as the Frank-Condon (FC) factors for the transitions. 
Because of the very low average laser intensity we are in a perturbative regime. Thus, we could  assume that the excitation probability is simply proportional to the laser spectral density at the transition frequencies, to the FC factor, and to the H\"onl-London factor. If needed rate equations can be performed, and exactly solved using for instance Kinetic Monte Carlo modeling~\cite{2008NJPh...10d5031C}, but in our strong perturbative regime, where much less than one photon is absorbed during the excited state lifetime ($\sim 15\,$ns), we simply assume that the all population has decayed before sending, 12.5~ns later, another broadband light pulse. 
	Initially, in each simulation, the molecules lie in the levels $v_{\rm X}=$1 to 7, with a distribution measured experimentally in Ref. \cite{2008Sci...321..232V} and corresponding to the  spectrum (without the femtosecond laser) of Fig.~\ref{fig2:v=0}d.

	In the first experiment, presented in Fig.~\ref{fig2:v=0}, we use our SLM setup in order to
 recover the results already presented in Ref. \cite{2008Sci...321..232V} where the shaping was simply realized by using a razor blade in the Fourier plane of a 4f-line.
 Starting from the sample of cold molecules in  vibrational levels $v_{\rm X}=$1 to 7 of the ground state, the idea is to use the  broadband laser to excite all populated vibrational levels but frequency-limited in such a way to eliminate  transitions from $v=0$ level, in which molecules accumulate.
For the cesium dimer, the frequencies that correspond to excitation of the ground vibrational level $v_{\rm X}=0$ to any vibrational level of the $B$ potential  lie higher than a specific threshold of ~13000 cm$^{-1}$. 
Consequently, the required laser shape
spectrum is  simply the usual laser spectrum truncated at this threshold.  The theoretical (assuming a gaussian shape and a 3\% extinction ratio) as well as the experimentally realized spectrum 
are presented in Fig.~\ref{fig2:v=0}b.
During the femtosecond excitation step, only a part of the transitions, between  the $v_{\rm X}$ and  $v_{\rm B}$ vibrational levels
of the ground $X $  and the excited $B$ electronic states,
 can occur since the available laser frequencies are limited. In order to understand the basics of the optical  pumping process we 
 represent (in hatched red) in Fig.~\ref{fig2:v=0}a  the transitions that can be excited  by the pulse shaped laser and
 (in grey)
  the Frank-Condon (FC) factors for the transitions 
  between the $v_{\rm X}$ and the $v_{\rm B}$ vibrational states. 
     The (grey) Frank-Condon parabola is useful to study the spontaneous decay of an excited molecule whereas the hatched red part is useful to study the laser excitation.
     The sharp energy cut in the laser spectrum is represented by a  solid line in this $v_{\rm X},v_{\rm B}$ graph because the energy varies almost linearly with the vibrational quantum number  due to the small anharmonicity in each of the X and B potential curves.    
   As example, let's follow the optical pumping of a molecule initially in the $v_{\rm X}=4$ level: its most probable
optical pumping walk is first to be excited into $v_{\rm B}=1$ (stronger FC factor in the hatched area) then to decay in
$v_{\rm X}=0$ (stronger FC factor) where no more excitation is possible (no hatched transition).
More generally, the result is given by the complete simulation and is shown in Fig.~\ref{fig2:v=0}c. After the application of 10$^4$ pulses, 70\% of the initial population, spread among several vibrational levels,  have been transferred into the sole $v_{\rm X}=0$ level. The remaining $30\%$ of the population is transferred to high vibrational levels that are not affected by the femtosecond laser because the possible transitions lie out of the range of frequencies available in the laser pulse we use. With a larger bandwidth laser, the simulation shows that a better efficiency of the process is possible (see Sect.~\ref{broadband}). 

When applying the shaped femtosecond laser to the cold molecules,
the experimental result is  shown in Fig.~\ref{fig2:v=0}d where
the detection of the vibrational level populated is done via a 2-photon REMPI scheme at 627 nm (DCM dye laser) via the $C^{1}\Pi _{u}$ state. We clearly see strong lines appearing at energies corresponding to $v_{\rm X} = 0 \rightarrow v_{\rm C} = 0, 1$ transitions, that are missing when the shaped femtosecond pulses are not applied.  
Due to some instabilities of
 the detection scheme, it is difficult to quantify the fraction of transfered population  which is theoretically of $70\%$.

	\subsection{Selective cooling to a single vibrational level}

\begin{figure}[htbp]
\begin{center}
\includegraphics[angle=-90,width=0.85\textwidth,keepaspectratio=true,clip=true,bb=11mm 16mm 200mm 270mm]{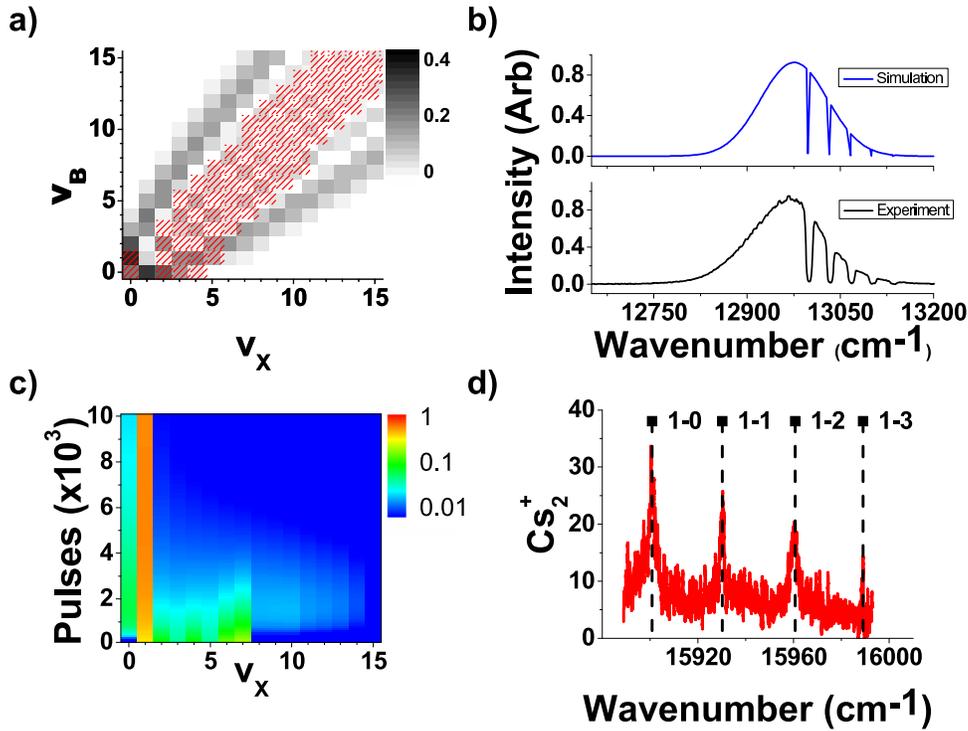}
\caption{Simulation and experimental results for the transfer in the $v_{\rm X} = 1$ level. a) FC coefficients (grey) for transition between the X ground state and the B excited state. The hatched (red) area represent transition allowed (i.e. having an intensity higher than 10\% of the maximum) by the shaped laser pulse: transitions from $v_{\rm X} = 1$ level, which is now the "target" state are completely removed.  b) The shaped pulse used for the simulation (upper part) and in the experiment (lower part). c) Results of the simulation of the vibrational cooling. The population of the $v_{\rm X}=1$ level after $10^4$ pulses is 53\%. d) Experimental detection spectrum of  molecule (mainly in $v_{\rm X}=1$) via two photon REMPI.}
\label{fig3:v=1}
\end{center}
\end{figure}

	The idea of removing the frequencies that correspond to (all) possible excitations of a particular vibrational level, in order to form a dark state where molecular population can accumulate with optical pumping, can be applied not only to the $v_{\rm X} = 0$ level. 
	Fig.~\ref{fig3:v=1} shows the case where the target vibration state is $v_{\rm X} = 1$ where we have shaped the pulse by removing any transition frequencies
between $v_{\rm X} = 1$ and $B$ state.
 The required spectrum and its realization by the pulse shaper are shown on Fig.~\ref{fig3:v=1}b.
	 Several tests have been performed in order to study the effect  of the number of pixels (between 1 and 5) used to make the "holes". No
substantial difference has been shown on the REMPI signal. In all the experiment two neighboring pixels are usually set to zero.
 This emphasizes that although our pulse
shaper has a high resolution, the molecular transitions are narrow enough to be
killed by only one "dark pixel". The $0.06\,$nm limited resolution of the pulse shaper does not cause any problem  as long as this does not lead to a second "dark" state in the system, a condition which is easily fulfilled due to the relatively large spectral separation ($\sim 40$cm$^{-1}$, corresponding to 2.3nm at 13000cm$^{-1}$) of the lower vibrational levels.
	 In Fig. \ref{fig3:v=1}b bellow, the experimental spectrum is recorded by a spectrometer (Ocean optics HR 4000) with a resolution twice less than the pulse shaper's one.

	  The simulation, shown in Fig.~\ref{fig3:v=1}c, predicts a total transfer of 
	  53\% to the $v_{\rm X} = 1$ level, value which could be increased up to
	  67\% if a perfect off-on ratio is simulated. Finally,  Fig.~\ref{fig3:v=1}d shows the detected ion spectrum, where frequencies corresponding to $v_{\rm X} = 1 \rightarrow v_{\rm C} = 0, 1, 2 , 3$ transitions, appear with a strong signal.

\begin{figure}[htbp]
\begin{center}
\includegraphics[width=15cm]{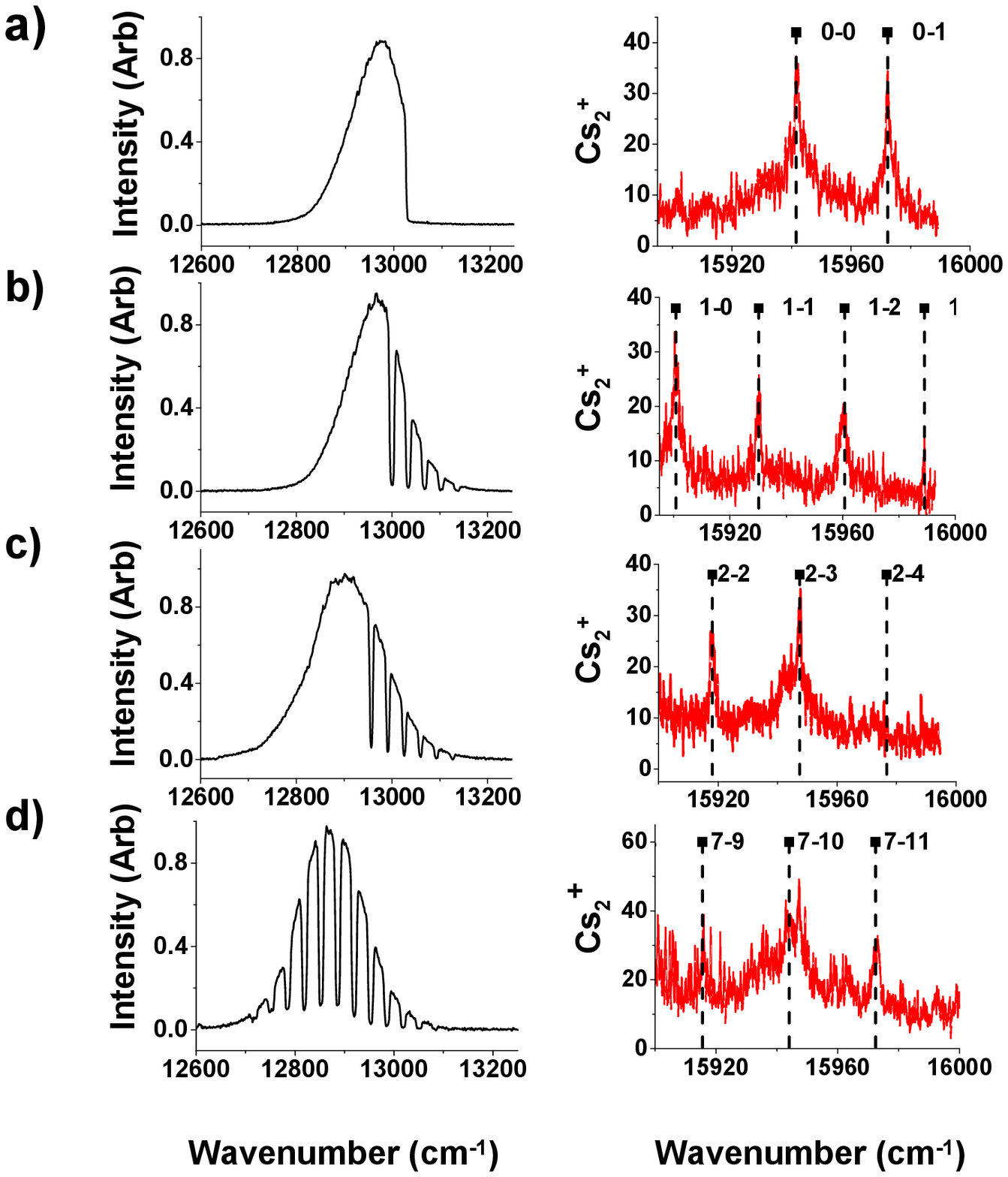}
\caption{Left part the experimental pulse spectra used to populate the $v_{\rm X}=0$ (a), $v_{\rm X}=1$ (b), $v_{\rm X}=2$ (c) and $v_{\rm X}=7$ (d) levels. Right part: the corresponding experimental ionization spectra. Notice that in spectrum (c)  a small signal corresponding to molecules in the $v_{\rm X}=0$ remains, and that in spectrum (d)  molecules in the $v_{\rm X}=1$ level also remain.}
\label{fig4:all_v}
\end{center}
\end{figure}

	The generality of the method is demonstrated in Fig.~\ref{fig4:all_v}, where the case for the $v_{\rm X}$= 0,1, 2 and 7 target states
	are presented.
	 For each chosen target $v_{\rm X}$,  the ionization spectrum contains mainly lines in positions corresponding to transitions from the selected target state to several excited vibrational levels. 
	
	In principle, any vibrational level can be chosen as the target state. The obvious limitation lies upon the available laser bandwidth and upon the initial molecular distribution. The laser has to be strong enough in the vicinity of transitions between the initial states and the target one. For the femtosecond laser used here (bandwidth ~55 cm$^{-1})$,  $v_{\rm X} = 7$ is an extreme choice, a fact that is indicated by a lower signal to noise ratio and the existence of $v_{\rm X} = 0, 1$ contributions to the signal. 

\subsection{Better shaping and accumulation analysis}

It has been demonstrated that population can be transferred to a desired vibrational level when the frequencies that connect it to any of the excited states used in the optical pumping scheme, are removed from the femtosecond pulse.  However,
the efficiency of such optical pumping procedure depends on the Frank-Condon factor, i.e. of the relative position of the electronically excited potential with respect to the ground-state one, on the bandwidth of the femtosecond pulse used and on the extinction ratio of the pulse shaper. 

We would like here to address  more complex pulse shaping which could possibly lead to a more efficient vibrational cooling, in terms of number of molecules finally transferred to the desired state,  than the one just described. As a particular example we will study the case of a "comb" of selected laser frequencies  chosen in such a way to induce only the transition required to produce efficient optical pumping from the initially populated levels to the target one. 
Several approaches  exist for the choice of such an optimized spectrum but of course, the target state has to remain a dark state and all frequencies resonant to it must be removed. 
A possible criterion for the choice of the allowed excited states is that their Franck-Condon coefficients with  the target vibrational level should be as high as possible. However,
in our case, due to the limited laser bandwidth,
it is more important to  limit 
the transfer  of  population  to high vibrational levels that are no more affected by the femtosecond laser.
Therefore, one has chosen
to favor excitation in levels that correspond to the "lower branch" of the FC parabola. With such a choice, once a $v_{\rm X}$ molecule is excited in this "lower branch", it will decay either in the same $v_{\rm X}$ level or in the "upper branch" of the FC parabola i.e. in lower vibrational levels than its initial value.

\begin{figure}[htbp]
\begin{center}
\includegraphics[angle=-90,width=0.85\textwidth,keepaspectratio=true,clip=true,bb=11mm 16mm 200mm 282mm]{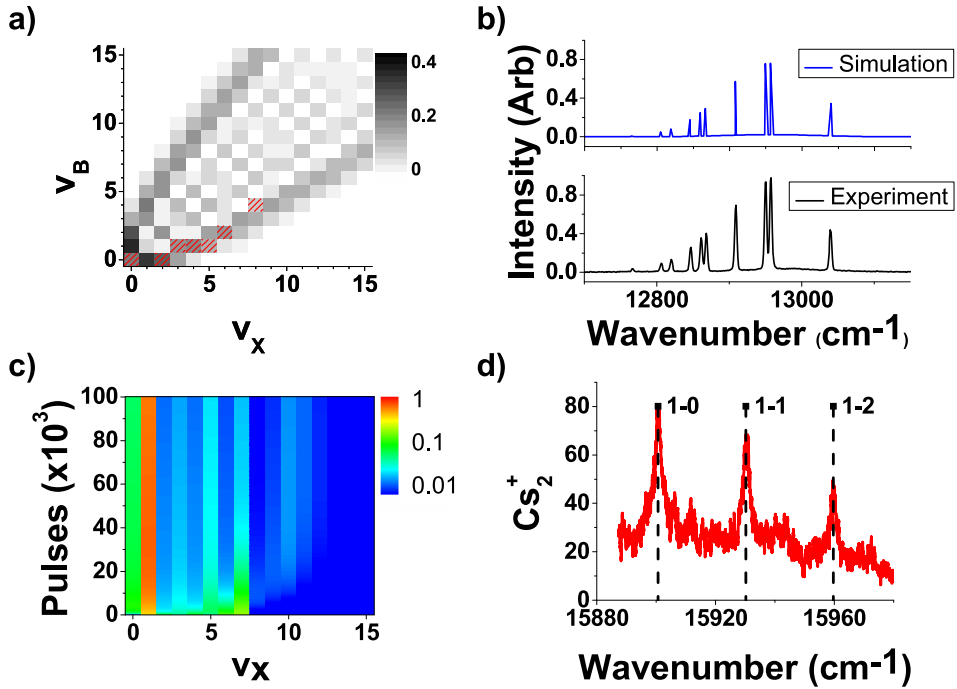}
\caption{Same as Figure \ref{fig3:v=1} but for the transfer in the $v_{\rm X} = 1$ level with an improved shaping. a) FC coefficients (grey) for transition between the X ground state and the B excited state. The hatched (red) area represents transitions allowed by the shaped laser pulse: i.e. between $v_{\rm X} \neq 1$  to $v_{\rm B}$ levels  that  decay  in levels
$v^{'}_{\rm X}
\leq v_{\rm X}$. b) The shaped pulse used for the simulation (upper part) and in the experiment (lower part). c) Results of the simulation of the vibrational cooling. The population of the $v_{\rm X}=1$ level after $10^5$ pulses is 57\% value which could be increased up to
	  98\% if a perfect off-on ratio is simulated. d) 
Experimental detection of (mainly $v_{\rm X}=1$) molecules via two photon REMPI.
}
\label{fig5:comb}
\end{center}
\end{figure}

An example is given  in Figure~\ref{fig5:comb} where the target state is again the $v_{\rm X} = 1$ level. Using the SLM  all laser frequencies 
are removed from the pulse spectrum,
except those that excite the various vibrational levels $v_{\rm X} \neq 1$  to levels $v_{\rm B}$ that  decay (see Eq. (\ref{equ:Equ1})) in levels
$v^{'}_{\rm X}
\leq v_{\rm X}$.

 The simulation of the required pulse (shown in Figure~\ref{fig5:comb}b upper part) shown in Figure~\ref{fig5:comb}c predicts a transfer for the "comb" pulse of 57\%, which is better than the 53\% of the corresponding "hole" pulse used in Figure~\ref{fig3:v=1}c. On the experimental side, the larger signal in Figure~\ref{fig5:comb}d with respect to the corresponding Figure~\ref{fig3:v=1}d seems to indicate that the population transfer is indeed, in the "comb" pulse case, more efficient  than using the "hole" pulse one. 
  
\begin{figure}[htbp]
\begin{center}
\includegraphics[width=15cm]{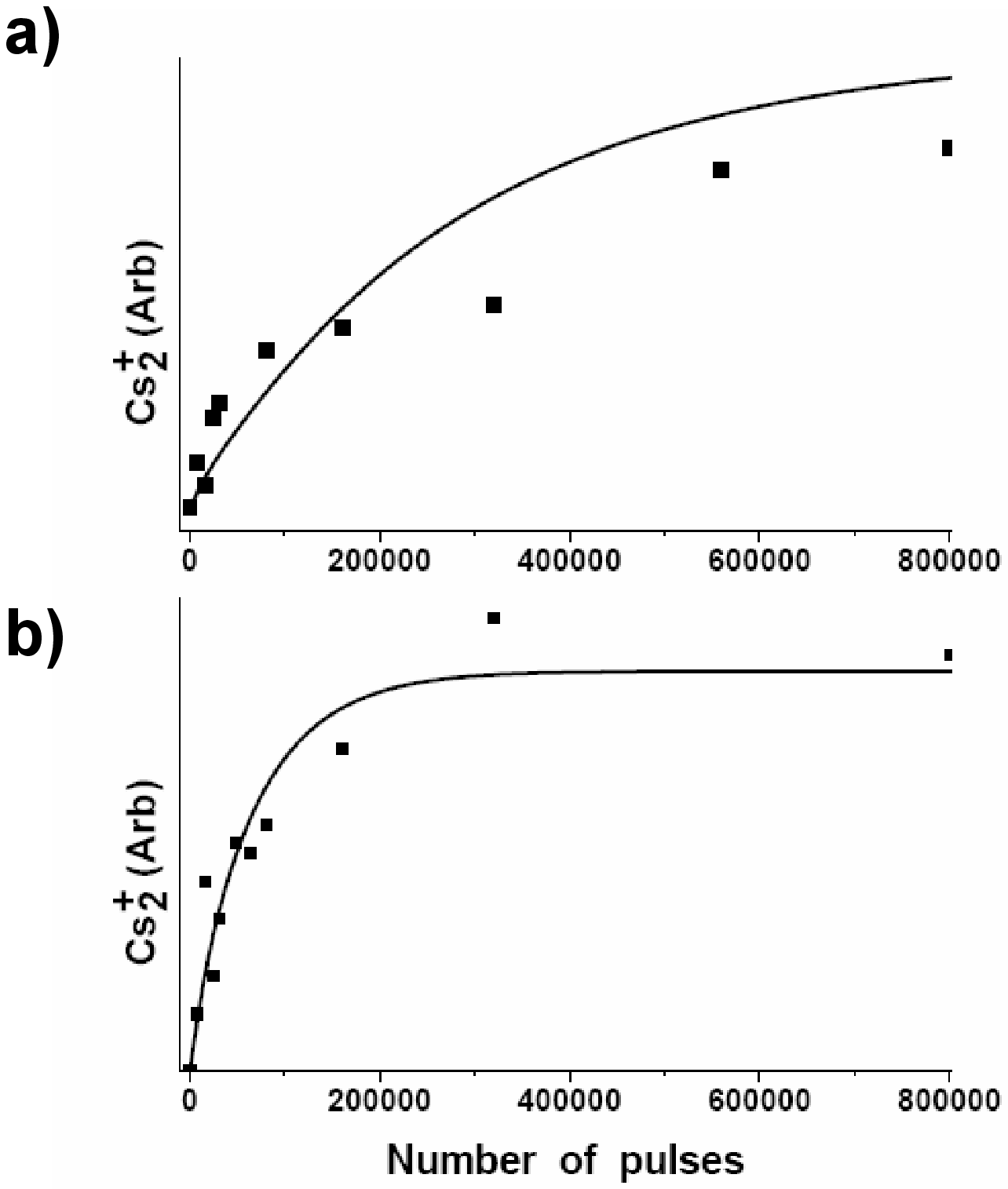}
\caption{Temporal evolution of the population of the $v_{\rm X}=1$ level. a) with the pulse plotted in Figure~\ref{fig5:comb}b and b) with the pulse plotted in Figure~\ref{fig3:v=1}b. The 
  lines correspond to the simulation and the dots to experimental measurements, i.e. to the Cs$_2^+$ ion signal recorded with the REMPI laser tuned to the transition $v_{\rm X}=1$ to $v_{\rm B}=0$ at $15900\,$cm$^{-1}$.}
\label{fig6:temporal}
\end{center}
\end{figure}

  By controlling the number of femtosecond laser pulses with an acousto-optic modulator, we analyzed the time dependence of the optical pumping scheme in
 Figure \ref{fig6:temporal}. One has recorded the temporal evolution of the vibrational cooling (as a function of the number of femtosecond pulses) with the use of the pulses shown in figures 3 and \ref{fig5:comb}, compared to the result of the simulation. Here again the behavior is the expected one. If it is difficult in the current experiment to make precise statement concerning the efficiency of the process, it is clear that, as expected, the optimized "comb" pulse is slower than the "hole" one." This is not a general feature but is simply due in our case to the limited laser bandwidth which limit the ability to excite the 
 "lower branch" of the FC parabola especially for high vibrational levels.

\section{Outlook and perspectives for a broadband laser cooling}
\label{broadband}

\begin{figure}[htbp]
\begin{center}
\includegraphics[angle=-90,width=0.85\textwidth,keepaspectratio=true,clip=true,bb=11mm 16mm 200mm 270mm]{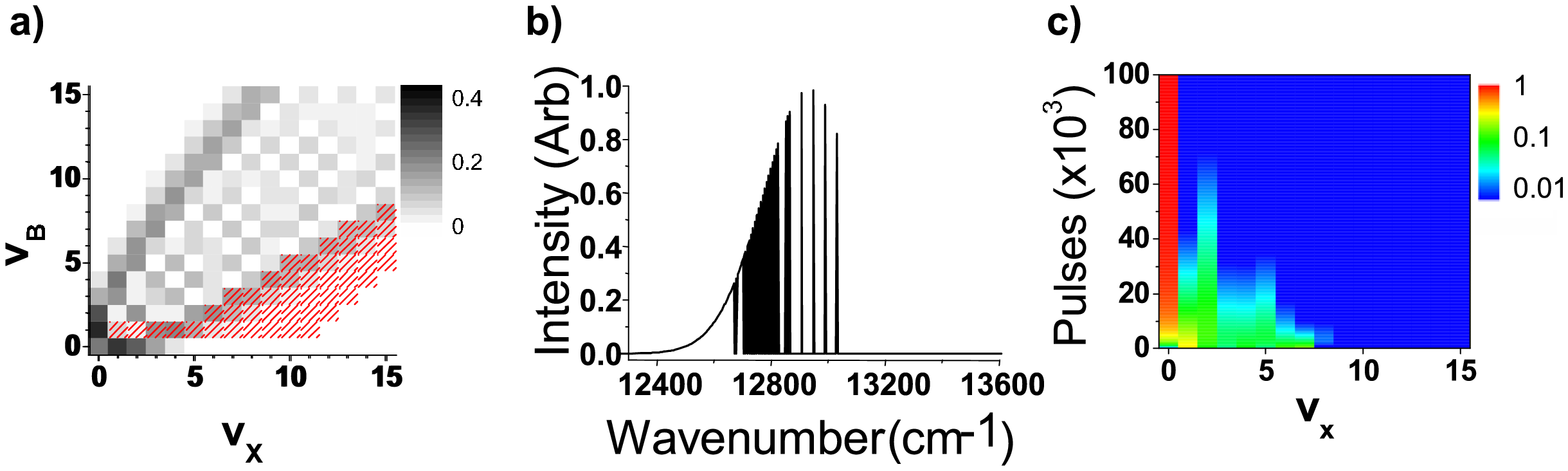}
\caption{Simulation for the vibrational cooling, similar to Figure \ref{fig5:comb}, but with the use of a broader shaped pulse (but with the same total intensity) and with a perfect off-on extinction ratio. a) FC coefficients (grey) for transition between the X ground state and the B excited state. The hatched (red) area represent absorption allowed transitions.  b) The shaped pulse used for the simulation. The bandwidth is now three times larger than the one of the previous pulses (165 cm$^{-1})$ c) Results of the simulation of the vibrational cooling. The population of the $v_{\rm X}=0$ level after $10^5$ pulses is 99.4\%}
\label{fig7:large}
\end{center}
\end{figure}

\subsection{Efficient accumulation of population} 

Due to the limited laser bandwidth and the imperfect extinction ratio of our SLM, the transfer efficiency seems  to be limited to roughly 60\%. The simulation 
clearly indicate that both effects are important. Indeed,  a three times broader laser, shaped in a similar manner than the previous one (excitation on the lower part of the Condon parabola),
would lead to a transfer efficiency toward $v_{\rm X} =1$ of 67\%. Furthermore,
as indicated by Figure \ref{fig7:large}, a perfect off-on ratio would even lead to an almost perfect transfer efficiency (99.4\%).

\subsection{Rotational cooling}
Let us note, that for the Cs dimer, the resolution 
required in order to separate all rotational levels would be on the order of 
$0.01$~cm$^{-1}$ (corresponding to 0.0006 nm), which is difficult to be achieved experimentally and beyond the capabilities  of the present pulse shaper which has a $0.06\,$nm resolution.
However,
the simulation can  be used to answer one important question:
can the laser cooling of the vibration to be extended to the rotation of the
molecules? As done with vibration we could think that an optimized shaping or just a "cut" could realize a rotational cooling. However we could not totally control the absorption step by an arbitrary shaping  due to the selection rules  $\Delta J=0,\pm 1$ so a more detailed study is required.

We would like to demonstrate that  rotational cooling is indeed in principle feasible  by studying the simple possible case of population transfer toward the lowest ro-vibronic state of the Cs$_2$ molecules namely $v_X=J_X=0$  level  where $J_{\rm X}$  represent the rotational quantum numbers in the X state.

\begin{figure}[htbp]
\begin{center}
\includegraphics[width=15cm]{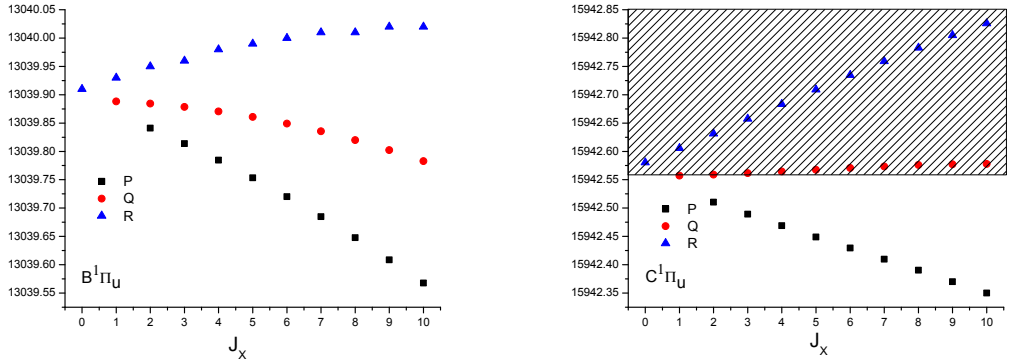}
\caption{Energy of the different ro-vibrationnel transitions (for the $\Delta J=0,\pm1$ Q R and P branches) (a) $\Delta J=J_{\rm B}-J_{\rm X}$
with the $B$ state, (b) $\Delta J=J_{\rm C}-J_{\rm X}$ with the $C$ state. The $C$ state permits rotational cooling with a simple laser shaping, which is not possible via the $B$ state due to the energy dependence of the transitions. By removing all frequencies larger than $15942.557$~cm$^{-1}$ we ensure that $\Delta J = J_{\rm C}-J_{\rm X}=-1$ in each excitation step except for the $J_{X}=1$ to $J_{C}=1$ transition. This way the $J_{X}=0$ is the only dark state of the system. }
\label{fig8:rot_en}
\end{center}
\end{figure}

A possibility for laser cooling of the molecular rotation
is to shape the laser by removing the frequencies corresponding to the
transitions $\Delta J=J_{\rm B}-J_{\rm X}=0,+1$, where $J_{\rm X}$ and $J_{\rm B}$ represent the rotational quantum numbers in each level. With such a shaping, 
absorption- spontaneous emission cycles would indeed lead to a decrease on average of
the principal rotation quantum number $J_{\rm X}$, \textit{i.e.} to a laser cooling
of the rotation. 
Figure~\ref{fig8:rot_en} shows the energies corresponding to
the rotational transitions between the vibrational levels $v_{\rm X}=0$ and 
$v_{\rm B}=0$. 

By analogy of what has been performed for the vibrational cooling realized in Figure \ref{fig2:v=0}  a simple "cut" in the laser spectrum could be implemented.
By a blue cut of the laser, it is not possible to suppress
the transitions $\Delta J=J_{\rm B}-J_{\rm X}=0$ (Q branch). This indicate that this simple "cut" of the spectrum would not be an efficient way to perform the rotational cooling  through the $B$ state. However,
rotational cooling through the $B$ state is possible by selecting only the $P$-branch using a shaper with a very high selectivity. 

But for 
simplicity we shall study another situation where
 we consider no longer the state, $B^{1}\Pi _{u}$, but the
state, $C^{1}\Pi _{u}$. Figure~\ref{fig8:rot_en} shows that the 
transitions $\Delta J=J_{\rm C}-J_{\rm X}=0$ can be easily suppressed by a simple energy cut.
Furthermore the high Franck-Condon value between $v_X=0$ and $v_C=0$ ensures that no spurious heating could occur by population transfer to $v_X \neq 0$ levels.

\begin{figure}[htbp]
\begin{center}
\includegraphics[width=15cm]{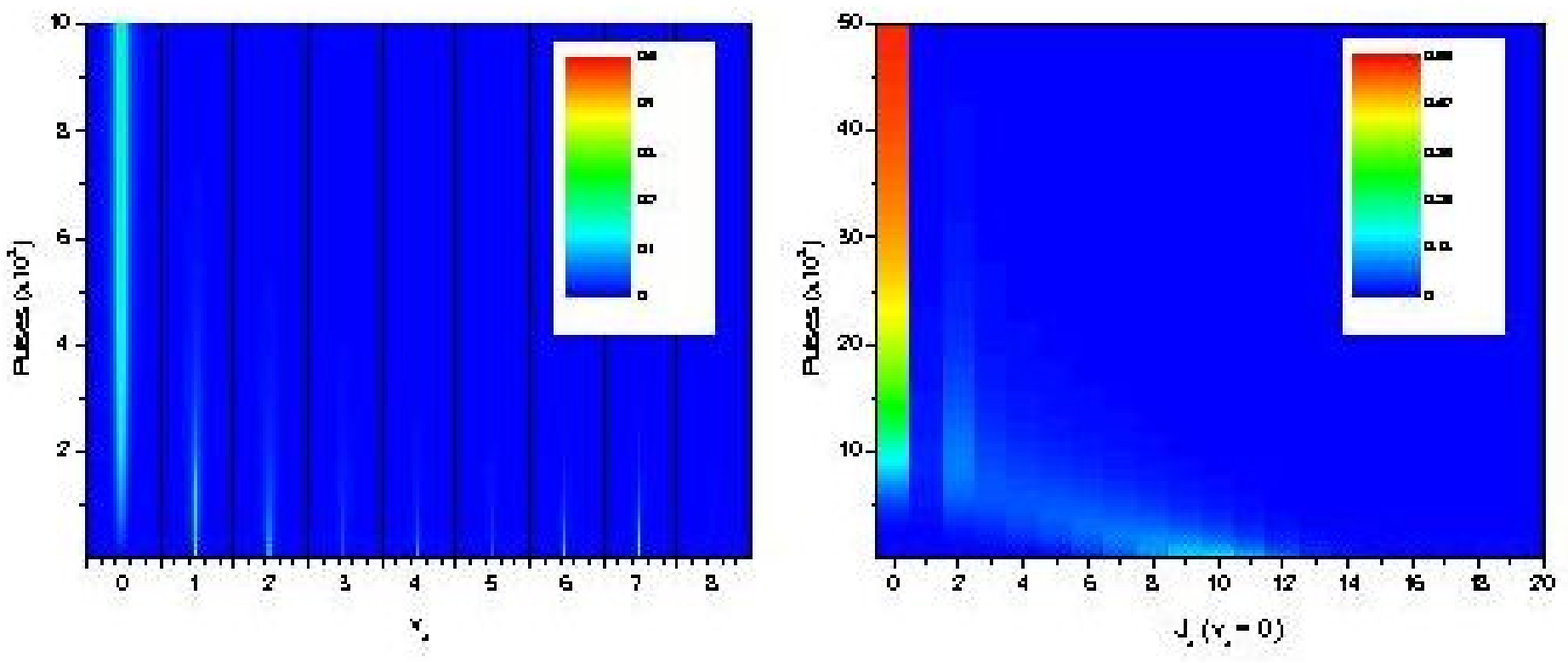}
\caption{Simulation of the temporal evolution for a ro-vibrational cooling. On the left side, the vibrational cooling to the  $v_X=0\,$ level, where the cooling is realized as in Figure \ref{fig2:v=0} via the state $B$ with all frequencies above $13000$~cm$^{-1}$ suppressed (just below the $ v_X=1\-\,v_B=0$ transition energy). Here the nine first vibrational levels are shown, each of them containing (shown from left to right in order) 21 rotational levels. Initially all population is placed in the $J=10$ level, a fact that does not affect the validity of the calculation, since the population is redistributed in the rotational levels under the influence of the optical pumping laser. On the right side, the rotational cooling to the $v_X=0\-\,J_X=0$, that follows this vibrational cooling, is realized via the state $C$ (laser wavelength $15940.0$~cm$^{-1}$, FWHM laser linewidth $50$~cm$^{-1}$), where all frequencies above $15942.557$~cm$^{-1}$ are suppressed, and with an initial rotational distribution which corresponds to the output of the vibrational cooling step.}
\label{fig9:rot}
\end{center}
\end{figure}

Figure~\ref{fig9:rot} shows the results of a simulation where the molecules
are first vibrationally cooled,  as done in Figure \ref{fig2:v=0} i.e. by applying the excitation of the state, $B^{1}\Pi _{u}$, then rotationally cooled by considering the excitation of the state $C^{1}\Pi _{u}$. Obviously such a scheme would require an external broadband source to excite the X states toward the C one, one could think of a  simple broadband diode laser  near $627\,$nm.

\section{Conclusion}

We have studied experimentally how
  femtosecond pulse shaping techniques can be used to realize efficient optical pumping of the vibration of cold Cs$_2$ molecules.
  We have used only a small part of the possibilities offered by the pulse shaping techniques, namely using it as an intensity spectral modulator for incoherent optical pumping. Using shaped laser with higher power could open several possibilities demonstrated to coherently transfer population between ro-vibrational levels \cite{1993JChPh..99..196B,1999FaraDisc,2001CP....267..195B,2001PhRvA..63a3407S}.
  However,  
the method demonstrated here is based on a light spectrally broad enough to excite large number of populated vibrational levels and shaped in amplitude such that it eliminates all frequencies that could excite the desire target state $v$. 
With an appropriate pulse shaper, we have demonstrated
the optical pumping to  a single vibrational levels such as $v=0,1,2$ and $v=7$. We have also demonstrated the possibility to optimize the pumping method by using convenient pulse shaping in order to excite only the most appropriate vibrational transitions. 
Rotational cooling can in principle be performed in a similar way provided
that the laser bandwidth and the experimental ability to shape the laser
matches the rotational energy spread.

The efficiency of the optical pumping is mainly
 limited by the finite laser spectral bandwidth and the imperfect extinction ratio of our SLM.
However
the theory indicates 
that the use of  broader sources and better off-on ratio 
have
 the possibility to accumulate near 100\% of population in one single vibrational level. Therefore the use
 of super-continuum source
 or a  simple broadband diode laser in combination with better (no gap) SLM or with mechanical shutters  might be interesting in this purpose. 
  This
opens the way to use such  a source as repumping light in a scheme for
 laser cooling of molecules~\cite{1996JChPh.104.9689B,2004EPJD...31..395D}.
Finally,
 accumulation of the molecules in an optical trap could lead 
to study of collisional processes in order to
assess the efficiency of evaporative cooling or to investigate ways for
achieving controlled chemistry.

\section{Acknowledgments}
This work is supported by the "Institut Francilien de Recherche sur les
Atomes Froids" (IFRAF) and (in Toulouse) by the Agence Nationale de la Recherche (Contract ANR
- 06-BLAN-0004) and the Del Duca foundation.

M.A. thanks the EC-Network EMALI. 
We thank Nadia Bouloufa and Olivier Dulieu for providing us with the FC calculations, and
Matthieu Viteau and Amodsen Chotia for the previous realization of part of the experimental setup.

\section{Bibliography}

\bibliographystyle{unsrt}
\bibliography{Shaping_NJP}

\begin{thebibliography}{10}

\bibitem{2008Sci...321..232V}
M.~{Viteau}, A.~{Chotia}, M.~{Allegrini}, N.~{Bouloufa}, O.~{Dulieu},
  D.~{Comparat}, and P.~{Pillet}.
\newblock {Optical Pumping and Vibrational Cooling of Molecules}.
\newblock {\em Science}, 321:232--234, July 2008.

\bibitem{Ye2005}
J.~Ye and S.T. Cundiff.
\newblock {\em Femtosecond optical frequency comb technology: principle,
  operation, and applications}.
\newblock Springer, NY, 2005.

\bibitem{Shapiro2003}
M.~Shapiro and P.~Brumer.
\newblock {\em Principles of the Quantum Control of Molecular Processes}.
\newblock Wiley-Interscience, Hoboken, NJ, 2003.

\bibitem{Alessandro2007}
D.~D'Alessandro.
\newblock {\em Introduction to Quantum Control and Dynamics}.
\newblock Chapman and Hall,Boca Raton, 2007.

\bibitem{2000Sci...288..824R}
H.~{Rabitz}, R.~{de Vivie-Riedle}, M.~{Motzkus}, and K.~{Kompa}.
\newblock {Whither the Future of Controlling Quantum Phenomena?}
\newblock {\em Science}, 288:824--828, May 2000.

\bibitem{2004EPJD...31..149D}
J.~{Doyle}, B.~{Friedrich}, R.~V. {Krems}, and F.~{Masnou-Seeuws}.
\newblock {Editorial: Quo vadis, cold molecules?}
\newblock {\em Eur. Phys. J. D}, 31:149--164, November 2004.

\bibitem{Krems}
Roman~V Krems.
\newblock {Molecules near absolute zero and external field control of atomic
  and molecular dynamics}.
\newblock {\em Int. Rev. Phys. Chem.}, 24:99 -- 118, 2005.

\bibitem{hutson-2006-25}
Jeremy~M. Hutson and Pavel Soldan.
\newblock Molecule formation in ultracold atomic gases.
\newblock {\em Int. Rev. Phys. Chem.}, 25:497, 2006.

\bibitem{DulieuJPB2006}
O.~{Dulieu}, M.~{Raoult}, and E.~{Tiemann}.
\newblock Cold molecules: a chemistry kitchen for physicists?
\newblock {\em J. Phys. B}, 39(19), Oct 2006.

\bibitem{2008PCCP...10.4079K}
R.~V. {Krems}.
\newblock {Cold controlled chemistry}.
\newblock {\em Physical Chemistry Chemical Physics (Incorporating Faraday
  Transactions)}, 10:4079--+, 2008.

\bibitem{Smith2008}
Ian W.~M. Smith.
\newblock {\em Low Temperatures and Cold Molecules}.
\newblock Imperial College Press, 2008.

\bibitem{2007PhRvL..99g3001M}
G.~{Morigi}, P.~W.~H. {Pinkse}, M.~{Kowalewski}, and R.~{de Vivie-Riedle}.
\newblock {Cavity Cooling of Internal Molecular Motion}.
\newblock {\em \prl}, 99(7):073001--+, August 2007.

\bibitem{1993JChPh..99..196B}
A.~{Bartana}, R.~{Kosloff}, and D.~J. {Tannor}.
\newblock {Laser cooling of molecular internal degrees of freedom by a series
  of shaped pulses}.
\newblock {\em \jcp}, 99:196--210, July 1993.

\bibitem{1999FaraDisc}
D.~J. {Tannor}, R.~{Kosloff}, and A.~{Bartana}.
\newblock {Laser cooling of internal degrees of freedom of molecules by
  dynamically trapped states}.
\newblock {\em Faraday Discuss.}, 113:365--383, 1999.

\bibitem{2001CP....267..195B}
A.~{Bartana}, R.~{Kosloff}, and D.~J. {Tannor}.
\newblock {Laser cooling of molecules by dynamically trapped states}.
\newblock {\em Chemical Physics}, 267:195--207, June 2001.

\bibitem{2001PhRvA..63a3407S}
S.~G. {Schirmer}.
\newblock {Laser cooling of internal molecular degrees of freedom for
  vibrationally hot molecules}.
\newblock {\em \pra}, 63(1):013407--+, December 2000.

\bibitem{2002PhRvL..89q3003V}
I.~S. {Vogelius}, L.~B. {Madsen}, and M.~{Drewsen}.
\newblock {Blackbody-Radiation-Assisted Laser Cooling of Molecular Ions}.
\newblock {\em \prl}, 89(17):173003--+, October 2002.

\bibitem{2004JPhB...37.4571V}
I.~S. {Vogelius}, L.~B. {Madsen}, and M.~{Drewsen}.
\newblock {Rotational cooling of molecules using lamps}.
\newblock {\em J. Phys. B}, 37:4571--4574, November 2004.

\bibitem{Assion98}
A.~Assion, T.~Baumert, M.~Bergt, T.~Brixner, B.~Kiefer, V.~Seyfried,
  M.~Strehle, and G.~Gerber.
\newblock Control of chemical reactions by feedback-optimized phase-shaped
  femtosecond laser pulses.
\newblock {\em Science}, 282:919, 1998.

\bibitem{MonmayrantCT-spirograph-06}
Antoine Monmayrant, Béatrice Chatel, and Bertrand Girard.
\newblock Quantum state measurement using coherent transients.
\newblock {\em Phys. Rev. Lett.}, 96(10):103002, 2006.

\bibitem{Motzkus02bio}
J.~L. Herek, W.~Wohlleben, R.~J. Cogdell, D.~Zeidler, and M.~Motzkus.
\newblock Quantum control of energy flow in light harvesting.
\newblock {\em Nature}, 417(6888):533, 2002.

\bibitem{efimovGA2000}
A.~Efimov, M.~D. Moores, B.~Mei, J.~L. Krause, C.~W. Siders, and D.~H. Reitze.
\newblock Minimization of dispersion in an ultrafast chirped pulse amplifier
  using adaptive learning.
\newblock {\em Appl. Phys. B}, 70(June):S133, 2000.

\bibitem{weinerjlt98}
H.~P. Sardesai, C.~C. Chang, and A.~M. Weiner.
\newblock A femtosecond code-division multiple-access communication system test
  bed.
\newblock {\em J. Lightw. Technol.}, 16(11):1953, 1998.

\bibitem{2000RScI...71.1929W}
A.~M. {Weiner}.
\newblock {Femtosecond pulse shaping using spatial light modulators}.
\newblock {\em Review of Scientific Instruments}, 71:1929--1960, May 2000.

\bibitem{Martinez87}
O.~E. Martinez.
\newblock 3000 times grating compressor with positive group velocity
  dispersion: application to fiber compensation in 1.3-1.6 mu m region.
\newblock {\em IEEE J. Quantum Electron.}, QE-23(1):59, 1987.

\bibitem{pulseshaperRSI04}
Antoine Monmayrant and Béatrice Chatel.
\newblock A new phase and amplitude high resolution pulse shaper.
\newblock {\em Rev. Sci. Instr.}, 75:2668, 2004.

\bibitem{1982JChPh..76.4370R}
M.~{Raab}, G.~{H{\"o}ning}, W.~{Demtr{\"o}der}, and C.~R. {Vidal}.
\newblock {High resolution laser spectroscopy of Cs$_{2}$. II. Doppler-free
  polarization spectroscopy of the C $^{1}\Pi _{u} \rightarrow X ^{1}Sigma ^{ +
  }_{g}$ system}.
\newblock {\em \jcp}, 76:4370--4386, May 1982.

\bibitem{2008PRA}
M.~{Viteau}, A.~{Chotia}, M.~{Allegrini}, N.~{Bouloufa}, O.~{Dulieu},
  D.~{Comparat}, and P.~{Pillet}.
\newblock {Efficient formation of deeply bound ultracold molecules probed by
  broadband detection}.
\newblock {\em \pra}, Accepted Mon Dec 29, 2008.

\bibitem{WeferNelson95slm}
M.~M. Wefers and K.~A. Nelson.
\newblock Analysis of programmable ultrashort waveform generation using
  liquid-crystal spatial light-modulators.
\newblock {\em J. Opt. Soc. Am. B}, 12(7):1343, 1995.

\bibitem{stobrawashap01}
G.~Stobrawa, M.~Hacker, T.~Feurer, D.~Zeidler, M.~Motzkus, and F.~Reichel.
\newblock A new high-resolution femtosecond pulse shaper.
\newblock {\em Appl. Phys. B}, 72(5):627, 2001.

\bibitem{1985JChPh..82.5354W}
W.~{Weickenmeier}, U.~{Diemer}, M.~{Wahl}, M.~{Raab}, W.~{Demtr{\"o}der}, and
  W.~{M{\"u}ller}.
\newblock {Accurate ground state potential of Cs$_{2}$ up to the dissociation
  limit}.
\newblock {\em J.\ Chem.\ Phys.}, 82:5354--5363, June 1985.

\bibitem{1989CPL...164..419D}
U.~{Diemer}, R.~{Duchowicz}, M.~{Ertel}, E.~{Mehdizadeh}, and
  W.~{Demtr{\"o}der}.
\newblock {Doppler-free polarization spectroscopy of the B $^{1}_{u}$ state of
  Cs$_{2}$}.
\newblock {\em \cpl}, 164:419--426, December 1989.

\bibitem{2008NJPh...10d5031C}
A.~{Chotia}, M.~{Viteau}, T.~{Vogt}, D.~{Comparat}, and P.~{Pillet}.
\newblock {Kinetic Monte Carlo modeling of dipole blockade in Rydberg
  excitation experiment}.
\newblock {\em New Journal of Physics}, 10(4):045031--+, April 2008.

\bibitem{1996JChPh.104.9689B}
J.~T. {Bahns}, W.~C. {Stwalley}, and P.~L. {Gould}.
\newblock {Laser cooling of molecules: A sequential scheme for rotation,
  translation, and vibration}.
\newblock {\em \jcp}, 104:9689--9697, June 1996.

\bibitem{2004EPJD...31..395D}
M.~D. {di Rosa}.
\newblock {Laser-cooling molecules}.
\newblock {\em European Physical Journal D}, 31:395--402, November 2004.

\end{thebibliography}

\end{document}